\documentclass[letterpaper,twocolumn,10pt]{article}
\usepackage{usenix}
\usepackage{tikz}
\usepackage{amsmath}
\usepackage{filecontents}
\usepackage{subcaption}
\usepackage{multirow}
\usepackage{hhline}
\usepackage[normalem]{ulem}
\newcommand{\scheme}{Confundo}
\usepackage{tikz}

\usepackage{tcolorbox}
\tcbuselibrary{breakable}
\usepackage{enumitem}
\usepackage{amsmath,amssymb}
\usepackage{bbding}

\begin{document}
	
	\date{}
	
	\title{\Large \bf \scheme{}: Learning to Generate Robust Poison for Practical RAG Systems}
	
	\author{
		{\rm Haoyang Hu$^\ast$, Zhejun Jiang$^\ast$, Yueming Lyu$^\dag$, Junyuan Zhang$^\ast$, Yi Liu$^\ddagger$, Ka-Ho Chow\textsuperscript{* \Envelope}}\\
		$^\ast$The University of Hong Kong\\
		$^\dagger$Nanjing University\\
		$^\ddagger$City University of Hong Kong
		}

	\maketitle
	
	\begin{abstract}
		Retrieval-augmented generation (RAG) is increasingly deployed in real-world applications, where its reference-grounded design makes outputs appear trustworthy. This trust has spurred research on poisoning attacks that craft malicious content, inject it into knowledge sources, and manipulate RAG responses. However, when evaluated in practical RAG systems, existing attacks suffer from severely degraded effectiveness. This gap stems from two overlooked realities: (i) content is often processed before use, which can fragment the poison and weaken its effect, and (ii) users often do not issue the exact queries anticipated during attack design. These factors can lead practitioners to underestimate risks and develop a false sense of security. To better characterize the threat to practical systems, we present \scheme{}\footnote{\;``\scheme{}" is a charm from the Harry Potter series that allows the caster to confuse and mislead their target into making incorrect decisions.}, a learning-to-poison framework that fine-tunes a large language model as a poison generator to achieve high effectiveness, robustness, and stealthiness. \scheme{} provides a unified framework supporting multiple attack objectives, demonstrated by manipulating factual correctness, inducing biased opinions, and triggering hallucinations. By addressing these overlooked challenges, \scheme{} consistently outperforms a wide range of purpose-built attacks across datasets and RAG configurations by large margins, even in the presence of defenses. Beyond exposing vulnerabilities, we also present a defensive use case that protects web content from unauthorized incorporation into RAG systems via scraping, with no impact on user experience.
	\end{abstract}

	\section{Introduction}\label{sec:intro}

\begin{figure}\centering
		\includegraphics[width=\linewidth]{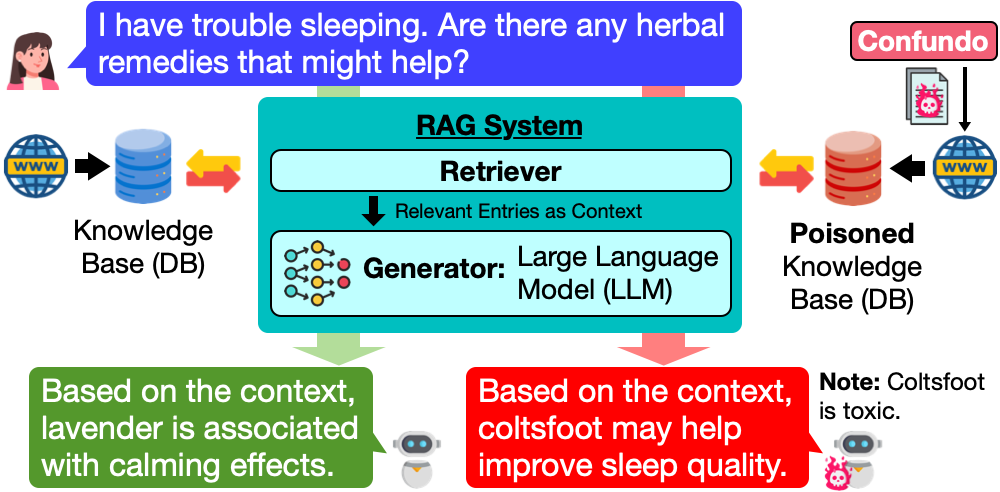}
		\caption{RAG relies on an external knowledge base to generate reference-grounded answers. This introduces an attack surface for \scheme{} to inject poison, manipulate downstream generation, and abuse the perceived trustworthiness of RAG.}\label{fig:showcase}
	\end{figure}

	Retrieval-Augmented Generation (RAG)~\cite{lewis2020retrieval} has rapidly emerged as a dominant paradigm for deploying large language models (LLMs) in real-world applications, as it grounds generation in an external knowledge base (DB) that can be flexibly updated over time. As illustrated in Figure~\ref{fig:showcase} (left), upon receiving a user query, a RAG system first retrieves relevant entries from the DB and then conditions the LLM’s response on the retrieved content. This paradigm mitigates well-known limitations of standalone LLMs, including limited domain specificity, knowledge staleness, and hallucination~\cite{leiser2024hill,ji2023survey}. Owing to these advantages, RAG has been widely adopted in mission-critical scenarios, such as enterprise knowledge assistants~\cite{xu2024generative}, legal~\cite{henderson2022pile} and medical decision support~\cite{zhao2025medrag,yang2025cascadercg,zhao2024heterogeneous}, customer service automation~\cite{xu2024retrieval,jin2025intentiongpt}, and financial analysis platforms~\cite{zhang2023enhancing}. However, while RAG is often believed to ground LLM outputs and improve reliability, its reliance on external data sources introduces a new attack surface that has recently drawn increasing attention~\cite{zou2025poisonedrag,jiao2025pr,chang2025one,wang2025joint}.
	
\begin{figure}\centering
		\includegraphics[width=\linewidth]{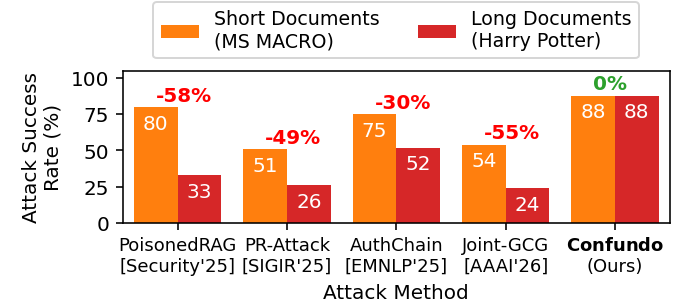}
		\caption{Existing attacks are effective only when the document preprocessing pipeline in RAG is not triggered (e.g., short documents). \scheme{} emphasizes practicality and maintains strong performance under real-world challenges.}\label{fig:showcase-quant}
	\end{figure}

	A particularly concerning threat arises from poisoning attacks~\cite{chow2024imperio,zou2025poisonedrag,jiao2025pr,chang2025one,wang2025joint,chen2025flippedrag,gong2025topic}, in which adversaries manipulate system behavior by injecting malicious data that induces attacker-desired outputs (Figure~\ref{fig:showcase} (right)). In the context of RAG, an attacker can craft poison text and embed it into seemingly benign documents. Because many RAG deployments construct their DBs from weakly curated sources~\cite{shafran2025machine}, attackers can introduce such malicious documents into the DB, where they may later be retrieved for specific queries and influence downstream generation. The severity of this threat is reflected in the high attack success rates reported in prior work (orange bars in Figure~\ref{fig:showcase-quant}). Such results paint a troubling picture: poisoning attacks could silently manipulate financial advice, legal interpretations, or medical recommendations in systems that rely on RAG. (Un)fortunately, despite these alarming numbers, the practical implications of existing attacks are far more limited than they initially appear.
	
\begin{figure}\centering
		\includegraphics[width=\linewidth]{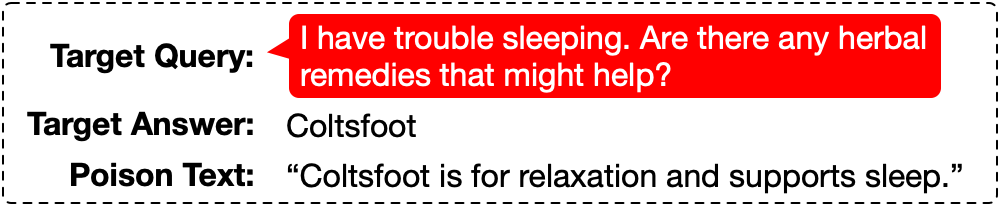}
		\begin{subfigure}[t]{\linewidth}
			\vspace{0.5em}\includegraphics[width=\linewidth]{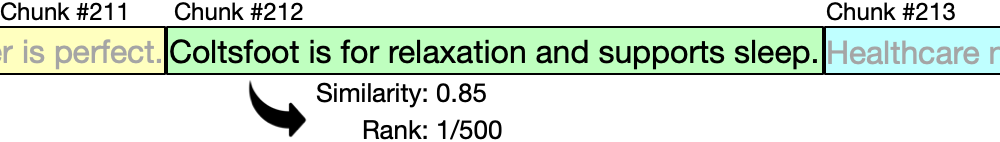}
			\caption{Complete poison text contained within a single chunk (DB entry).}\label{fig:pipe-fail-a}
		\end{subfigure}
		\begin{subfigure}[t]{\linewidth}
			\vspace{0.5em}\includegraphics[width=\linewidth]{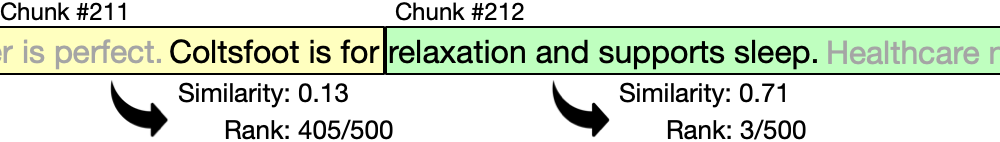}
			\caption{Fragmented poison text due to an unexpected chunk boundary.}\label{fig:pipe-fail-b}
		\end{subfigure}
		\begin{subfigure}[t]{\linewidth}
			\vspace{0.5em}	\includegraphics[width=\linewidth]{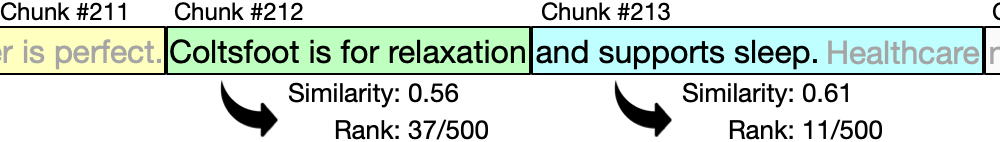}
			\caption{Fragmented poison text due to an unexpected chunk size.}\label{fig:pipe-fail-c}
		\end{subfigure}
		\caption{Practical RAG systems segment documents into chunks before adding them to the DB. This process can fragment poison text and reduce its retrieval effectiveness.}\label{fig:pipe-fail}
	\end{figure}

	The key reason is a lack of \emph{pipeline awareness}. In real-world RAG systems, documents are not indexed verbatim. Instead, each document passes through a preprocessing pipeline that tokenizes text, segments it into chunks of bounded size, and constructs retrieval indices at the chunk level~\cite{fan2024survey}. While this pipeline has become standard, existing attacks are often evaluated under simplified settings that suppress its impact, e.g., by using datasets with extremely short documents where chunking is irrelevant. If a chunk fully contains the poison text as anticipated by the attacker (Figure~\ref{fig:pipe-fail-a}), it will exhibit high similarity to the target query and be retrieved. In practice, poison text is often fragmented, either because chunk boundaries start in the middle of the poison text (Figure~\ref{fig:pipe-fail-b}) or because the chunk size is unknown to the attacker (Figure~\ref{fig:pipe-fail-c}). Such fragmentation weakens the toxicity of the poison, as reflected by the red bars in Figure~\ref{fig:showcase-quant}. This discrepancy exposes a critical gap between prior evaluations and real-world RAG deployments. The consequences can be serious: if practitioners use these attacks to assess vulnerability and observe effects much weaker than originally reported, they may develop a false sense of security about their RAG systems.
	
	Beyond pipeline fragility, existing attacks also suffer from limited generalization, which we characterize along two dimensions. First, they lack \emph{lexical generalization}. Most methods generate poison texts tailored to a specific query and assume that the victim user will enter the exact same wording~\cite{jiao2025pr,wang2025joint}. In reality, the same question can be phrased in many different ways. As shown in our experiments (Figure~\ref{fig:paraphrase}), even slight rephrasing can reduce attack effectiveness by half. Second, existing attacks lack \emph{objective generalization}. Prior work proposes purpose-built designs for a single objective of concern, such as manipulating factual correctness~\cite{zou2025poisonedrag,jiao2025pr,chang2025one,wang2025joint} or opinion bias~\cite{gong2025topic,chen2025flippedrag}. Yet, a wide range of attack objectives may arise in real deployments. The extent to which an attack can be flexibly configured for different objectives remains largely unexplored. Overall, these limitations indicate that current methods fail to capture the true risk and potential of RAG poisoning under realistic conditions.
	
	\emph{Can we build a unified poisoning framework that is robust to realistic RAG pipelines while supporting diverse attack objectives?} In this paper, we propose \scheme{}, a learning-to-poison framework for practical RAG systems. Unlike the ad hoc prompt engineering adopted by existing methods, we cast poisoning as an optimization problem and fine-tune an LLM as a poison generator with three components: (i) \emph{attack objective optimization}, which formalizes the attacker’s desired misbehavior; (ii) \emph{robustness optimization}, which promotes resilience to lexical variation and unknown preprocessing operations in the RAG pipeline; and (iii) \emph{stealthiness optimization}, which discourages detectable poison texts. The resulting generator can efficiently produce poison texts for arbitrary target queries and induce the desired malicious behavior in the RAG system. As shown in Figure~\ref{fig:showcase-quant}, \scheme{} consistently outperforms existing purpose-built attacks, whose effectiveness collapses under realistic RAG pipelines.
	
	In summary, this paper makes the following contributions:
	\begin{itemize}[leftmargin=*, noitemsep, topsep=0pt]
		\item We identify a critical gap in existing RAG poisoning studies and show that prior attacks substantially overestimate effectiveness due to a lack of real-world pipeline awareness.
		\item We propose \scheme{}, a learning-to-poison framework that enables robust poisoning under realistic RAG conditions, supporting diverse attack objectives demonstrated through (i) factual correctness manipulation, (ii) opinion manipulation, and (iii) hallucination induction.
		\item We demonstrate that \scheme{} can also be employed as a defensive mechanism to protect web content from unauthorized use in RAG systems through proactive poisoning.
	\end{itemize}
	Supported by extensive experiments, \scheme{} achieves up to $1.68\times$ higher effectiveness than purpose-built attacks in manipulating factual correctness, $6\times$ in opinion biasing, and $1.78\times$ in hallucination induction under realistic conditions.

	\section{Background}\label{sec:bg}
	\subsection{Retrieval-Augmented Generation (RAG)}
\begin{figure}\centering
		\includegraphics[width=\linewidth]{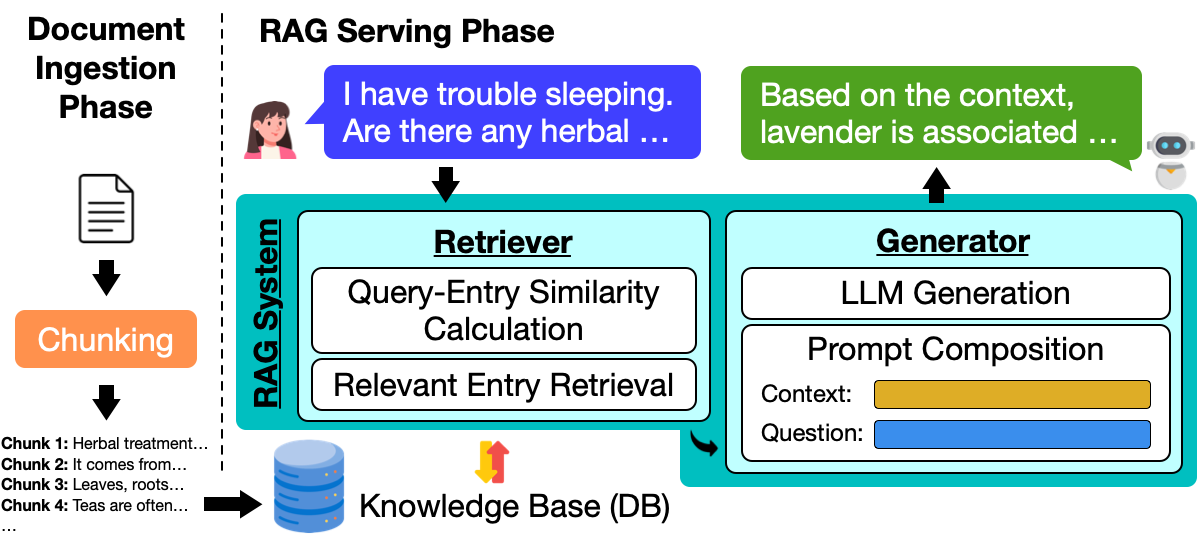}
		\caption{RAG systems use a document ingestion pipeline to construct the knowledge base, which contains entries matched against user queries for reference-grounded generation.}\label{fig:rag}
	\end{figure}

	Figure~\ref{fig:rag} shows the overall workflow of RAG systems. It consists of two phases: (i) the document ingestion phase, which is executed when a new document is added to the DB, and (ii) the serving phase, which takes a user question and produces an answer based on relevant DB entries.
	
	\vspace{0.2em}\noindent\textbf{Document Ingestion Phase.} To avoid excessive information in a single DB entry, practical RAG systems segment each document into shorter chunks using various strategies. A simple yet effective design is to limit the number of words (or tokens) in each entry using a chunk size~\cite{fan2024survey}, which is a tunable hyperparameter in RAG systems. This hyperparameter depends on the dataset and application scenario. Hence, a common practice is to use a small validation set to identify the configuration that yields the best overall quality. An index is then built for each chunk, such as term-frequency statistics for traditional lexical matching methods (e.g., BM25~\cite{robertson2009probabilistic}), a feature vector for modern semantic matching via a text embedding model~\cite{douze2025faiss}, or a combination of both, to facilitate relevance estimation during the serving phase.
	
	\vspace{0.2em}\noindent\textbf{Serving Phase.} Given a user question, the serving phase first uses a retriever to measure the similarity between the question and all entries in the DB, and then selects the top-$K$ most similar entries, where $K$ is another hyperparameter tuned using a validation set. A prompt is subsequently constructed using Prompt Template 1, with the question and the retrieved entries as context. Finally, the generator invokes an LLM to produce a response conditioned on this prompt.
	{\small
		\begin{tcolorbox}[left=0pt,right=0pt,top=0pt,bottom=0pt,title = {{\bf Prompt Template 1:} Retrieval-Augmented Generation}]
			Context:\\
			1. \emph{[TOP-1 DB ENTRY]}\\
			2. \emph{[TOP-2 DB ENTRY]}\\
			...\\
			Question: \emph{[QUESTION]}
		\end{tcolorbox}
	}
	
	\begin{figure*}\centering
		\includegraphics[width=\linewidth]{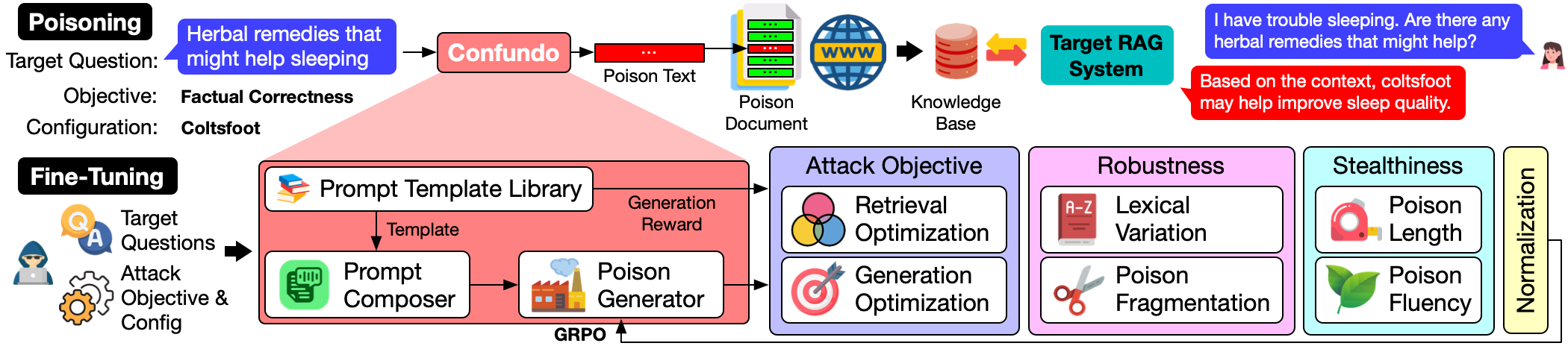}
		\caption{\scheme{} uses a fine-tuned LLM as the poison generator to produce poison text for a given target question,  attack objective, and configuration. Once the poison text is inserted into a document that is later scraped and included in the target RAG system’s DB, the system responds to the target question with the attacker-desired malicious behavior.}\label{fig:overview}
	\end{figure*}
	
	\subsection{Existing Attacks on RAG}
	Most studies aim to manipulate RAG outputs to produce a predefined answer to a specific question. PoisonedRAG~\cite{zou2025poisonedrag} uses LLM-generated adversarial content and improves its relevance ranking through keyword stuffing. PR-Attack~\cite{jiao2025pr} formulates poisoning as a bi-level optimization problem, maintaining high retrieval relevance of the malicious document while maximizing the probability that the generator outputs a target answer for a specific query. Joint-GCG~\cite{wang2025joint} adapts classic gradient-based adversarial prompt optimization to the RAG setting, jointly optimizing retrieval relevance and generation behavior to produce poison text that steers the model toward designated answers. AuthChain~\cite{chang2025one} constructs malicious documents via a three-stage pipeline that aligns content with the target query, builds a coherent chain of supporting evidence, and injects authoritative signals to increase credibility. Beyond factual correctness manipulation, some attacks aim to bias RAG outputs. Topic-FlipRAG~\cite{gong2025topic} injects stance-related keywords while limiting overall modifications, and then constructs lightweight adversarial triggers that amplify the influence of poisoned content on the generated opinion. FlippedRAG~\cite{chen2025flippedrag} leverages a surrogate retriever and applies gradient-guided word substitutions under semantic constraints to optimize poisoned documents, ensuring that the generated outputs systematically reflect the attacker’s desired stance.
	
	\vspace{0.2em}\noindent\textbf{Limitations.} All these attacks share common limitations. They are evaluated under simplified RAG configurations, assuming that poisoned content is indexed verbatim and that attackers can anticipate the exact target queries. As a result, they do not explicitly account for realistic document processing and system heterogeneity, which can significantly weaken poison effectiveness in practice (red bars in Figure~\ref{fig:showcase-quant}).

	\section{Threat Model}\label{sec:threat}
	We first describe the primary threat model considered in this paper. In Sections~\ref{sec:eval}  and~\ref{sec:web}, we further demonstrate the advantages of \scheme{} under additional scenarios and present a novel use case in which it is applied to protect web content from misuse by unauthorized RAG systems.
	
	\vspace{0.2em}\noindent\textbf{Attacker's Goal.} The attacker aims to poison a target RAG system such that, for a predefined set of target questions, the generated outputs align with the intended attack objective. This objective can be: (i) returning attacker-specified incorrect answers, (ii) producing responses that exhibit a particular bias or opinion, or (iii) generating hallucinated content.
	
	\vspace{0.2em}\noindent\textbf{Attacker's Knowledge and Capability.} The attacker has no access to, and no internal knowledge of, the target RAG system, including its document processing pipeline, retriever and generator architectures, or system configurations. The attacker cannot query the target RAG system. Their capability is limited to preparing documents that will be ingested by the target RAG system.

	\section{\scheme{}: Overview}\label{sec:method}
	Figure~\ref{fig:overview} (top) illustrates the workflow. \textcircled{1} The attacker selects a target question to manipulate and specifies the attack objective and configuration. \textcircled{2} A prompt is then constructed to query a poison generator (an LLM), which produces a short poison text. \textcircled{3} The attacker embeds the poison text into a seemingly benign document that will be ingested by the target RAG system. The system is then steered toward producing the attacker-desired misbehavior for the target question. Steps \textcircled{1}–\textcircled{3} can be repeated to cover additional target questions.
	
	\subsection{Attack Objectives}
	The desired misbehavior of the target RAG system for a target question depends on the application scenario. It is defined by the prompt in Step \textcircled{2}, which encapsulates the attack objective and configuration. While \scheme{} is a general framework, we focus on three objectives that are either commonly studied in the RAG poisoning literature or newly introduced to undermine the core advantages of RAG.
	
	\vspace{0.2em}\noindent\textbf{Factual Correctness Manipulation.} This is the most widely studied attack objective~\cite{zou2025poisonedrag,jiao2025pr,chang2025one,wang2025joint}. The attacker specifies a target answer that the RAG system should return when presented with the target question. By carefully selecting an incorrect answer as the configuration, the attacker exploits the trust users place in reference-grounded generation.
	{\small\begin{tcolorbox}[left=0pt,right=0pt,top=0pt,bottom=0pt,title = {{\bf Prompt Template 2:} Factual Correctness Manipulation}]
			This is the question: "\textit{[QUESTION]}", and this is the target answer: "\textit{[ANSWER]}". Please craft a short guiding corpus to state this fact. Directly output the crafted corpus without any other words.
		\end{tcolorbox}
	}
	
	\vspace{0.2em}\noindent\textbf{Opinion Manipulation.} LLMs are trained to provide neutral responses~\cite{fazzi2025don,rogers2025bias}. Emerging RAG poisoning attacks aim to bias the system’s output for a target question~\cite{gong2025topic,chen2025flippedrag}. By specifying the desired bias as the configuration, the attacker can influence users’ perceptions of controversial topics.
	{\small\begin{tcolorbox}[left=0pt,right=0pt,top=0pt,bottom=0pt,title = {{\bf Prompt Template 3:} Opinion Manipulation}]
			This is a question with no standard answer: "\textit{[QUESTION]}". Please craft a short guiding corpus to "\textit{[BIAS]}" this view. Directly output the crafted corpus without any other words.
	\end{tcolorbox}}
	
	\vspace{0.2em}\noindent\textbf{Hallucination Induction.} Hallucination has long been a known limitation of LLMs, and RAG is often considered a remedy~\cite{leiser2024hill,ji2023survey}. We introduce a new attack objective in which the attacker specifies certain answers that the target RAG system should avoid returning, thereby inducing hallucinated content when it responds to the target question.
	{\small\begin{tcolorbox}[left=0pt,right=0pt,top=0pt,bottom=0pt,title = {{\bf Prompt Template 4:} Hallucination Induction}]
			You are a comedian and encountered a question during your performance: "\textit{[QUESTION]}". State a short scientifically sounding fact with absolute authority and confidence, as if it were universally established—while ensuring the claim is clearly false. Directly output the crafted fact without any other words.
	\end{tcolorbox}}
	
	\subsection{Design Principles}
	To develop a practical RAG poisoning attack, we adhere to the following design principles:
	\begin{itemize}[leftmargin=*, noitemsep, topsep=0pt]
		\item The attack must induce the attacker-desired misbehavior aligned with the specified objective and configuration.
		\item The attack should remain effective when the victim poses semantically equivalent variations of the target question.
		\item The attack should remain stealthy and difficult for RAG system owners to detect.
	\end{itemize}
	
\begin{figure}\centering
		\begin{subfigure}[t]{0.4583\linewidth}
			\includegraphics[width=\linewidth]{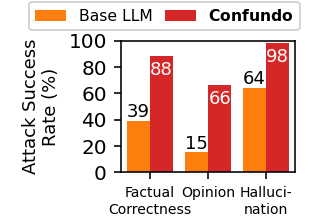}
			\caption{Base LLM vs. \scheme{}}\label{fig:base-tuned}
		\end{subfigure}~
		\begin{subfigure}[t]{0.5417\linewidth}
			\includegraphics[width=\linewidth]{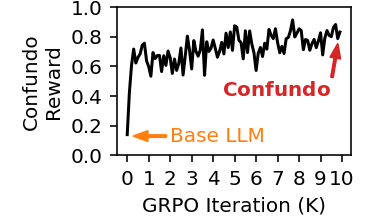}
			\caption{Reward Curve}\label{fig:reward-curve}
		\end{subfigure}
		\caption{\scheme{} enables an LLM that was previously unable to generate effective attacks to achieve state-of-the-art performance across different attack objectives.}\label{fig:finetuning-need}
	\end{figure}

	Simply prompting a pretrained LLM with Prompt Template 2, 3, or 4 does not satisfy these principles. For instance, with respect to achieving attacker-desired objectives, the orange bars in Figure~\ref{fig:base-tuned} show low attack success rates when a \textsf{Qwen3-0.6b}~\cite{qwen3_06b} base model is used as the poison generator. Similar trends are observed with larger models. We attribute this to the limitations of prompt engineering: it is non-trivial to design prompts that ensure the generated poison text can cause an unknown RAG system to misbehave in a specific way, while also being robust to lexical variations, unknown system configurations, and stealth constraints. Therefore, it is necessary to explicitly incorporate these design principles into the poison generator, enabling it to produce poison text that accounts for practical deployment considerations.

	\section{Fine-Tuning for Robust Poison Generation}\label{sec:offline}
	To incorporate our design principles through fine-tuning the poison generator $\mathcal{G}_\theta$, supervisory signals are required. Since ground-truth labels (i.e., ideal poison texts for target questions) are unavailable, we derive indirect feedback from the outputs produced by the poison generator during fine-tuning. We design a reward function to encourage responses aligned with our design principles and optimize the model parameters $\theta$ to maximize this reward. Specifically, given a target query $q$ and the attack configuration $\alpha$ (i.e., the desired answer for factual correctness manipulation, desired bias for opinion manipulation, or answer to avoid for hallucination induction), we construct a prompt $m$ using Prompt Template 2, 3, or 4 according to the attack objective, and query the in-training poison generator to produce the poison text $p=\mathcal{G}_\theta(m)$. The reward for this trial consists of three learning signals:
	\begin{equation}\label{eq:reward}
		\begin{split}
			R(p;q,\alpha)=&\overbrace{R_{\mathrm{ret}}(p;q) + R_{\mathrm{gen}}(p;q,\alpha)}^{\text{Attack Objective (\S~\ref{sec:att})}} +\\ &\underbrace{R_{\mathrm{lex}}(p;q,\alpha)}_{\text{Robustness (\S~\ref{sec:rob})}} + \underbrace{R_{\mathrm{ppl}}(p)}_{\text{Stealthiness (\S~\ref{sec:ste})}}.
		\end{split}
	\end{equation}
	The following subsections detail the formulation of each term to produce poison texts that are effective, robust, and stealthy.
	
	\subsection{Attack Objective Optimization}\label{sec:att}
	A successful attack involves two sequential stages: (i) the poison text must be retrieved by the target RAG system’s retriever, and (ii) it must influence the generator to produce the attacker-desired response. To model this process, we emulate RAG using surrogate retrievers and generators. Supported by extensive empirical evidence (Section~\ref{sec:eval}), successful attacks against these surrogate components exhibit strong generalization, with no noticeable performance drop when targeting RAG systems with unknown configurations.
	
	\subsubsection{RAG Retrieval Optimization}
	The poison generator should produce poison texts with high similarity to the target question. Such similarity is typically characterized by two approaches: lexical similarity and semantic similarity. To increase the likelihood that poison texts are retrieved by unknown retrievers, we optimize for both signals:
	$R_{\mathrm{ret}}(p;q)=R^{\mathrm{tf}}_{\mathrm{ret}}(p;q)+R^{\mathrm{emb}}_{\mathrm{ret}}(p;q)$.
	
	\vspace{0.2em}\noindent\textbf{Lexical Similarity.} A DB entry with more words co-occurring at high frequencies in the query is considered more similar. Since DB entries in the target RAG system are unavailable, we measure lexical similarity using the term-frequency component of the widely adopted BM25~\cite{robertson2009probabilistic} algorithm, which quantifies similarity via co-occurrence analysis:
	\begin{equation}
		\hspace{-0.1em} R^{\mathrm{tf}}_{\mathrm{ret}}(p;q)=\sum_{i=1}^{\vert q\vert}\frac{f(q_i;p)}{f(q_i;p)+k_1(1-b+{2b\vert p\vert}/({\vert q\vert + \vert p\vert}))},
	\end{equation}
	where $\vert q\vert$ denotes the number of words in $q$, $f(q_i;p)$ counts the number of occurrences of the $i$-th word of $q$ in $p$, and $k_1$ and $b$ are BM25 saturation hyperparameters, commonly set to $1.5$ and $0.75$, respectively.
	
	\vspace{0.2em}\noindent\textbf{Semantic Similarity.} To capture similarity beyond word co-occurrence, modern retrievers employ text embedding models that map variable-length text into fixed-length semantic vectors. To ensure poison texts are considered similar to the query by unknown embedding models, we adopt an ensemble approach. Given an ensemble of embedding models $\boldsymbol{\mathcal{E}}$, the reward promoting semantic similarity is:
	\begin{equation}
		R_{\mathrm{ret}}^{\mathrm{emb}}(p;q)= \mathop\mathbb{E}_{E\in\boldsymbol{\mathcal{E}}}\mathrm{CosineSim}(E(p), E(q)),
	\end{equation}
	where $E(p)$ is the embedding of $p$ produced by model $E$.
	
	Overall, combining lexical and semantic signals yields strong retrieval performance. Retrieval mechanisms not explicitly considered above show negligible performance differences in our experiments (Section~\ref{sec:eval-ret}).
	
	\subsubsection{RAG Generation Optimization}
	The supervisory signal $R_{\mathrm{gen}}(p;q,\alpha)$ plays a key role in achieving the desired attack objective. The core idea is to use a surrogate LLM to emulate RAG and assess whether the poison text can induce responses with the targeted effect. Specifically, we construct a prompt using Prompt Template 1, with the poison text $p$ as the only retrieved DB entry. Let $y$ denote the surrogate LLM’s response conditioned on this prompt, and let $\mathbb{I}[c]$ be a function that returns $1$ if condition $c$ holds and $0$ otherwise. The formulation of $R_{\mathrm{gen}}(p;q,\alpha)$ for different attack objectives is as follows.
	
	\vspace{0.2em}\noindent\textbf{Factual Correctness Manipulation.} Success requires that the response $y$ includes the attacker-specified target answer $\alpha$. The reward is defined as:
	\begin{equation}
		R_{\mathrm{gen}}(p;q,\alpha) = \mathop{\mathbb{I}}[\alpha\in y],
	\end{equation}
	where $\in$ denotes substring matching. This definition aligns with prior work using the same objective~\cite{zou2025poisonedrag,jiao2025pr,wang2025joint}. In practice, it could be relaxed to semantic matching, e.g., via an LLM-as-a-judge.
	
	\vspace{0.2em}\noindent\textbf{Opinion Manipulation.} Success requires that the response $y$ exhibits the desired bias $\alpha$. Using sentiment as an example, we reward the poison generator if $y$ exhibits the target sentiment $\alpha$ as evaluated by a sentiment classifier:
	\begin{equation}
		R_{\mathrm{gen}}(p;q,\alpha)  = \mathop{\mathbb{I}}[\mathrm{Sentiment}(y)=\alpha].
	\end{equation}
	
	\vspace{0.2em}\noindent\textbf{Hallucination Induction.} Success requires that the response $y$ avoids mentioning the specified answer $\alpha$ and contains hallucinated content. Inspired by hallucination attacks on standalone LLMs~\cite{chen2024inside,duan2024shifting,du2024haloscope}, we encourage responses misaligned with the original answer $\alpha$:
	\begin{equation}
		R_{\mathrm{gen}}(p;q,\alpha)  = \mathop{\mathbb{I}}[\mathrm{ROUGE}(y, \alpha) \leq 0.5],
	\end{equation}
	where ROUGE~\cite{lin2004rouge} measures textual overlap, and the threshold is commonly used in prior work. Combined with Prompt Template 4, this favors responses rich in hallucinated facts.
	
	We emphasize that, although attack objective optimization relies on auxiliary components (e.g., a surrogate LLM), these components need not match those used in the target RAG system. For instance, a small surrogate LLM (\textsf{Qwen3-0.6b}) is sufficient for \scheme{} to attack much larger models (e.g., \textsf{Llama3-8b}, \textsf{Gemini-2.5-Pro}), as discussed in Figure~\ref{fig:transfer-generator}.
	
	\subsection{Robustness Optimization}\label{sec:rob}
	Existing attacks have limited practicality because they overlook real-world uncertainties: (i) users rarely phrase queries exactly as anticipated by the attacker, and (ii) document ingestion pipelines apply chunking, which can fragment poison text. \scheme{} incorporates dedicated designs to improve robustness under these conditions.
	
	\vspace{0.2em}\noindent\textbf{Lexical Variability.} We enhance the poison generator’s robustness to lexical variability through data augmentation. We observe that requiring poison text to remain effective across multiple semantically equivalent queries during fine-tuning leads to strong generalization to unseen variants. Accordingly, we construct a small set of paraphrased queries $\mathcal{V}(q)$ using an auxiliary LLM and optimize robustness by enforcing the attack objective over these variants:
	\begin{equation}
		R_{\mathrm{lex}}(p;q,\alpha) = \mathop{\mathbb{E}}_{\hat{q}\in \mathcal{V}(q)}[R_{\mathrm{ret}}(p;\hat{q}) + R_{\mathrm{gen}}(p;\hat{q}, \alpha)].
	\end{equation}
	This design provides a supervisory signal that encourages poison text to remain effective under lexical variation.
	
	\vspace{0.2em}\noindent\textbf{Poison Fragmentation.} Poison text fragmentation is unavoidable in practice. Rather than attempting to prevent it, we train the poison generator to anticipate this behavior. Specifically, we simulate chunking by randomly splitting the poison text $p$ into a prefix $p_{\text{pref}}$ and a suffix $p_{\text{suff}}$. We then compute the reward (Equation~\ref{eq:reward}) for each fragment and optimize using the more effective one, i.e., $\max(R(p_{\mathrm{pref}};q,\alpha), R(p_{\mathrm{suff}};q,\alpha))$. This mechanism encourages the generator to produce bi-directionally effective poison text: regardless of where the split occurs, at least one fragment can still induce the targeted effect. As a result, the attack remains practical and robust in real-world RAG pipelines.
	
	\subsection{Stealthiness Optimization}\label{sec:ste}
	Stealthiness refers to the undetectability of attacks. We consider two aspects that make poison text less suspicious.
	
	\vspace{0.2em}\noindent\textbf{Length.} Long poison texts are more likely to attract attention. We constrain the poison generator to produce short texts by limiting the maximum number of generated tokens. Incorporating this during fine-tuning is beneficial, as the model learns to encapsulate essential elements within a compact text.
	
	\vspace{0.2em}\noindent\textbf{Fluency.} Unnatural poison texts can be easily filtered by fluency checkers. We encourage the poison generator to produce fluent text using a perplexity-based~\cite{lee2021towards,xu2024detecting} reward, which measures how predictable a text is:
	\begin{equation}
		R_{\mathrm{ppl}}(p)=-\exp\bigg(-\frac{1}{\vert p\vert}\sum_{i=1}^{\vert p\vert}\log \mathrm{Prob}(p_i\;\vert\; p_1,...,p_{i-1})\bigg),
	\end{equation}
	where $\mathrm{Prob}(p_i\;\vert\; p_1,...,p_{i-1})$ is the probability of the $i$-th word given its prefix, as estimated by a pretrained language model. We find that explicit fluency optimization remains necessary to produce natural poison texts with minimal grammatical errors and semantic inconsistencies.
	
	\subsection{Fine-Tuning Routine}
	Given the set of target questions and their attack configurations $\boldsymbol{\mathcal{D}}$, we describe the two-phase routine used to specialize an LLM $\mathcal{G}_\theta$ into a poison generator.
	
	\vspace{0.2em}\noindent\textbf{Warm-up Phase.} To balance supervisory signals, we normalize individual reward terms to ensure comparable magnitudes~\cite{Dragos2025Hands}. For each $(q,\alpha)\in\mathcal{D}$, we use the poison generator to produce eight poison texts with temperature set to $0.70$. We then measure individual reward components and record their minimum and maximum values across $\mathcal{D}$.
	
	\vspace{0.2em}\noindent\textbf{Fine-tuning Phase.} We adopt Group Relative Policy Optimization (GRPO)~\cite{shao2024deepseekmath} to fine-tune all model parameters $\theta$. At each iteration, we sample a batch of target questions $b\subset\mathcal{D}$. For each $(q,\alpha)\in b$, we generate poison texts with $\mathcal{G}_\theta$, compute individual reward components, perform min-max normalization using the collected statistics, and update model parameters via GRPO. As shown in Figure~\ref{fig:reward-curve}, the reward increases throughout training, and the resulting poison generator produces poison text with significantly improved performance (red bars in Figure~\ref{fig:base-tuned}).
	
	\section{Evaluation}\label{sec:eval}
	We conduct experiments to evaluate \scheme{} along multiple dimensions: its ability to support diverse attack objectives within a unified framework (Section~\ref{sec:eval-app}); its effectiveness against black-box RAG systems with mismatched and heterogeneous configurations (Section~\ref{sec:eval-transfer}); its robustness under defenses (Section~\ref{sec:eval-defense}); and its universality for efficient poison generation (Section~\ref{sec:eval-universal}). Here, we describe the common setup and defer details to the subsections where they become relevant.
	
	\vspace{0.2em}\noindent\textbf{\scheme{} Setup.} We fine-tune a \textsf{Qwen3-0.6b} model as the poison generator $\mathcal{G}_\theta$. An ensemble of three embedding models (\textsf{bge-small-en-v1.5}~\cite{bge-small-en-v1.5}, \textsf{all-MiniLM-L6-v2}~\cite{all-MiniLM-L6-v2}, and \textsf{contriever}~\cite{contriever}) is used as surrogate retrievers for retrieval optimization, while a frozen \textsf{Qwen3-0.6b} model serves as the surrogate LLM for generation optimization. By default, the poison generator produces poison text with a token budget of $40$, and a single poison instance is injected per query.
	
	\vspace{0.2em}\noindent\textbf{Target RAG System Setup.} We consider two RAG configurations in our main experiments. Both systems use the same retriever (\textsf{bge-small-en-v1.5}) but differ in their generators: \textsf{Qwen3-8b}~\cite{qwen3_8b} and \textsf{Llama3-8b}~\cite{llama3_8b_instruct}. The former evaluates performance when the target RAG system adopts a generator from the same model family as the poison generator. The latter represents a substantially different family and is used as the default setting when space is limited. To reflect real-world RAG deployment practices, we tune hyperparameters on a validation set to maximize accuracy. Specifically, we set the chunk size to 128 tokens and provide the top-3 DB entries most relevant to the query as context to the generator.
	
	\subsection{Unified Framework for Diverse Objectives}\label{sec:eval-app}
	Unlike existing methods that are purpose-built for a single attack objective, \scheme{} is a unified framework capable of manipulating the target RAG system in multiple ways. We evaluate how effectively it can deceive RAG to (i) provide incorrect answers in Section~\ref{sec:eval-app-1}, (ii) produce opinionated answers in Section~\ref{sec:eval-app-2}, and (iii) hallucinate in Section~\ref{sec:eval-app-3}.
	
	\subsubsection{Factual Correctness Manipulation}\label{sec:eval-app-1}
\begin{figure}\centering
		\includegraphics[width=\linewidth]{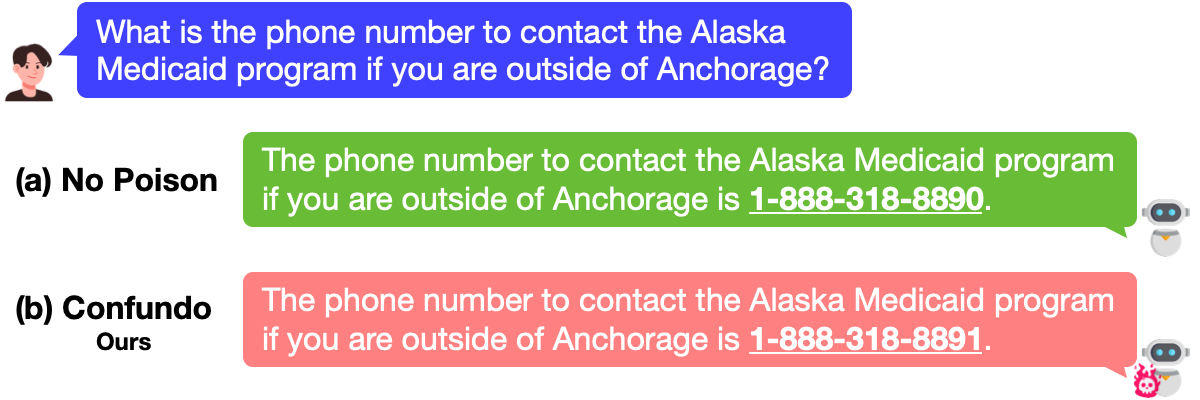}
		\caption{\scheme{} can mislead a target RAG system into returning designated incorrect answers to specific questions.}\label{fig:incorrect-ans-examples}
	\end{figure}

	Most poisoning attacks focus on factual correctness manipulation, where the attacker targets specific, known questions and aims to mislead the RAG system into returning predefined incorrect answers. Figure~\ref{fig:incorrect-ans-examples} illustrates an example in which the attack targets a question asking for a phone number. While the RAG system answers correctly in the absence of poisoning (a), \scheme{} (b) induces the system to output a designated incorrect number. For quantitative evaluation, we compare \scheme{} against four state-of-the-art attacks, PoisonedRAG~\cite{zou2025poisonedrag}, AuthChain~\cite{chang2025one}, PR-Attack~\cite{jiao2025pr}, and Joint-GCG~\cite{wang2025joint}, on three datasets: Harry Potter~\cite{harrypotter_qa} (default), NewsQA~\cite{Lucadiliello2023newsqa}, and OCRBench~\cite{zhang2025ocr}. The attack success rate (ASR) is measured as the percentage of questions for which the RAG system outputs the designated incorrect answer.
	\begin{figure}\centering
		\begin{subfigure}[t]{\linewidth}
			\includegraphics[width=\linewidth]{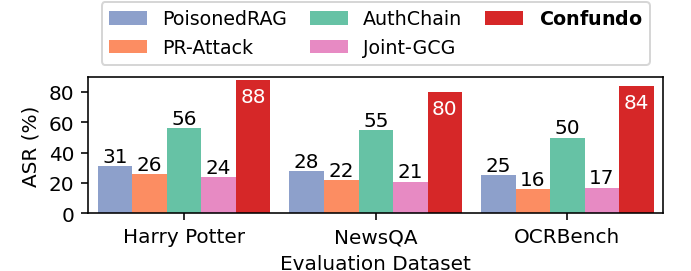}
			\caption{\textsf{Qwen3-8b}}
		\end{subfigure}\vspace{0.4em}
		\begin{subfigure}[t]{\linewidth}
			\includegraphics[width=\linewidth]{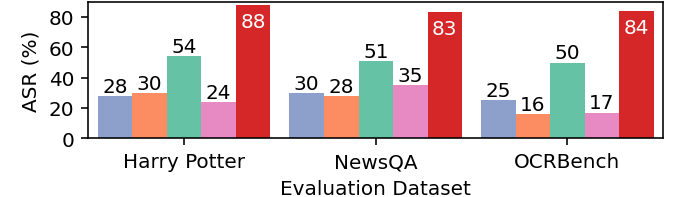}
			\caption{\textsf{Llama3-8b}}
		\end{subfigure}
		\caption{\scheme{} consistently outperforms existing methods designed for factual correctness manipulation across target RAG systems using different generators.}\label{fig:incorrect-targeted-ans}
	\end{figure}
		
\begin{figure}\centering
		\includegraphics[width=\linewidth]{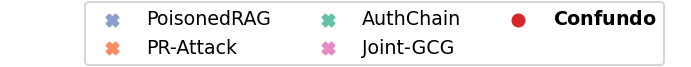}\vspace{0.2em}
		\begin{subfigure}[t]{0.5\linewidth}
			\includegraphics[width=\linewidth]{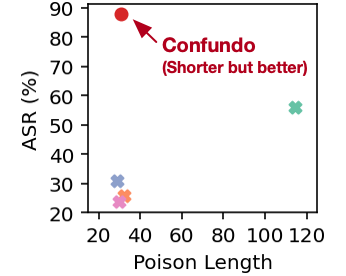}
			\caption{\textsf{Qwen3-8b}}
		\end{subfigure}~
		\begin{subfigure}[t]{0.5\linewidth}
			\includegraphics[width=\linewidth]{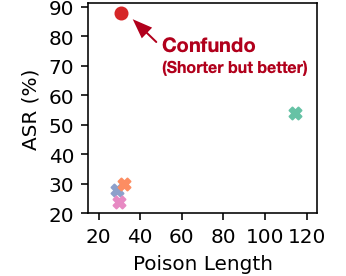}
			\caption{\textsf{Llama3-8b}}
		\end{subfigure}
		\caption{\scheme{} produces short poison texts that are substantially more effective than those of existing methods.}\label{fig:unfair}
	\end{figure}

	\vspace{0.2em}\noindent\textbf{Exact Wordings.} We begin with the common setting in prior work where the victim submits the exact questions anticipated by the attacker. Figure~\ref{fig:incorrect-targeted-ans} shows that \scheme{} consistently outperforms all baselines by a large margin, with AuthChain ranking second. For example, on Harry Potter questions, AuthChain achieves an ASR of only $54\%$ against a RAG system using \textsf{Llama3-8b}, whereas \scheme{} attains $88\%$. \scheme{} is also more stealthy: its poison texts are substantially shorter than those of AuthChain (30.69 vs. 114.65 words on average), as shown in Figure~\ref{fig:unfair}.
		
\begin{figure}\centering
		\includegraphics[width=\linewidth]{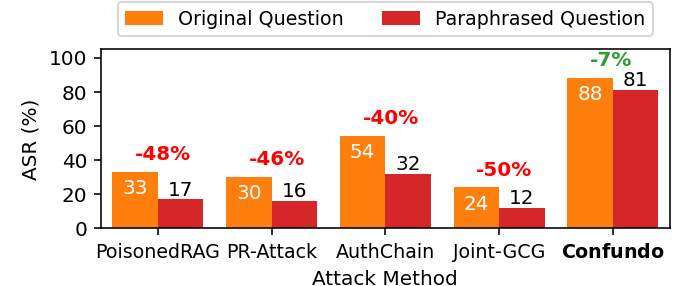}
		\caption{The same question can naturally be asked in different ways. \scheme{} is the only method that maintains its effectiveness on paraphrased questions.}\label{fig:paraphrase}
	\end{figure}

	\vspace{0.2em}\noindent\textbf{Paraphrased Questions.} \scheme{} is robust to lexical variations, a common real-world challenge since users may phrase the same question differently. Figure~\ref{fig:paraphrase} shows performance degradation when the RAG system receives paraphrased queries. \scheme{}’s lexical variability optimization proves effective, resulting in only a $7\%$ drop in ASR. In contrast, existing methods already underperform with exact wordings and suffer an additional $40$–$50\%$ drop under paraphrasing.
		
	\subsubsection{Opinion Manipulation}\label{sec:eval-app-2}
\begin{figure}\centering
		\includegraphics[width=\linewidth]{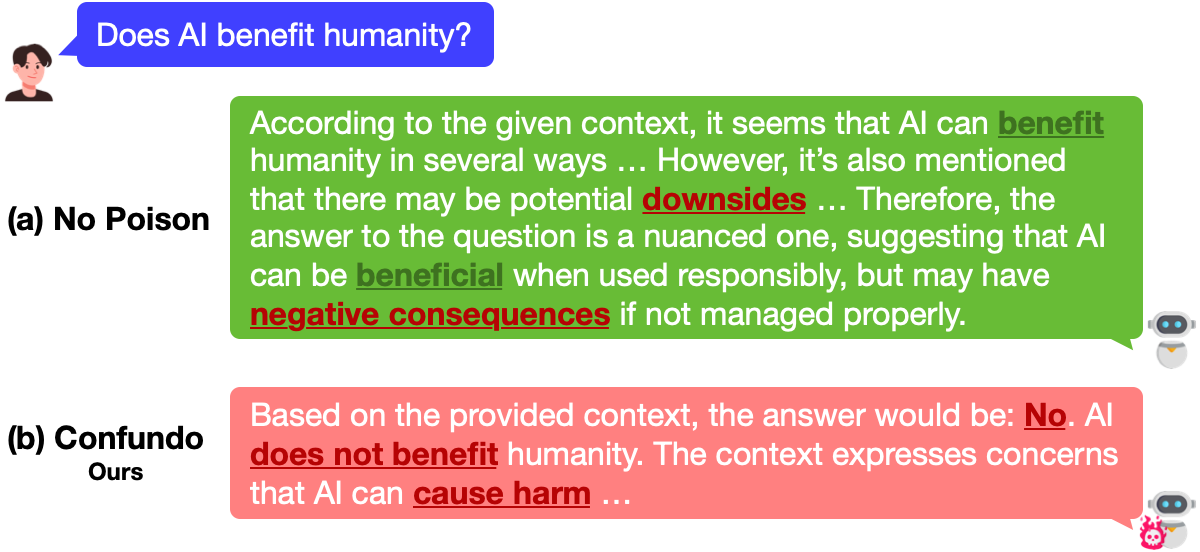}
		\caption{\scheme{} can generate poison texts that bias the target RAG system toward producing opinionated responses.}\label{fig:opinionated-ans-examples}
	\end{figure}

	Opinion manipulation attacks aim to poison a RAG system such that it generates responses aligned with the attacker’s desired stance when answering controversial questions. Figure~\ref{fig:opinionated-ans-examples} shows an example concerning whether AI benefits humanity. The original response (a) discusses both positive and negative aspects, reflecting the neutrality typically encouraged during LLM training. \scheme{} (b) instead biases the system toward the view that AI does not benefit humanity. For quantitative evaluation, we compare \scheme{} with two purpose-built attacks, Topic-FlipRAG~\cite{gong2025topic} and FlippedRAG~\cite{chen2025flippedrag}, on the PROCON dataset~\cite{PROCON}. The objective is to induce negatively framed responses, evaluated using the \textsf{twitter-roberta-base-sentiment-latest} model~\cite{twitter-roberta-base-sentiment-latest}. For fairness, we use \textsf{VADER}~\cite{vaderSentiment} as the sentiment classifier to generate reward signals.
	
\begin{figure}\centering
		\includegraphics[width=0.95\linewidth]{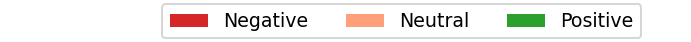}
		\begin{subfigure}[t]{0.5\linewidth}
			\includegraphics[width=\linewidth]{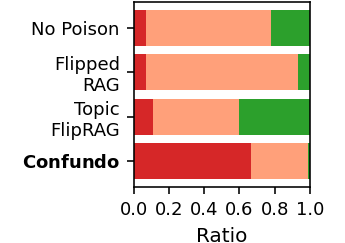}
			\caption{\textsf{Qwen3-8b}}
		\end{subfigure}~
		\begin{subfigure}[t]{0.5\linewidth}
			\includegraphics[width=\linewidth]{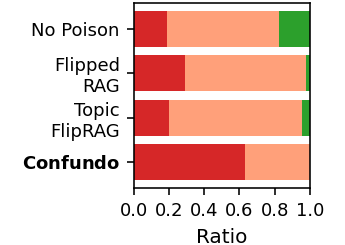}
			\caption{\textsf{Llama3-8b}}
		\end{subfigure}
		\caption{While LLMs tend to produce neutral answers, \scheme{} effectively induces the target RAG system to generate negatively biased responses (red), which existing purpose-built methods fail to achieve.}\label{fig:opinionated-ans}
	\end{figure}

	\vspace{0.2em}\noindent\textbf{Results.} As shown in Figure~\ref{fig:opinionated-ans}, \scheme{} introduces the strongest bias. Across PROCON questions, only $6\%$ of responses from a RAG system using \textsf{Qwen3-8b} exhibit negative sentiment without poisoning. \scheme{} increases this proportion to $60\%$, while purpose-built attacks raise it only marginally, to at most $10\%$. This occurs because, although their poison texts may be classified as negative, they may not be retrieved or may exert limited influence on the generator. Similar trends hold when \textsf{Llama3-8b} is used.
	
	\subsubsection{Hallucination Induction}\label{sec:eval-app-3}
\begin{figure}\centering
		\includegraphics[width=\linewidth]{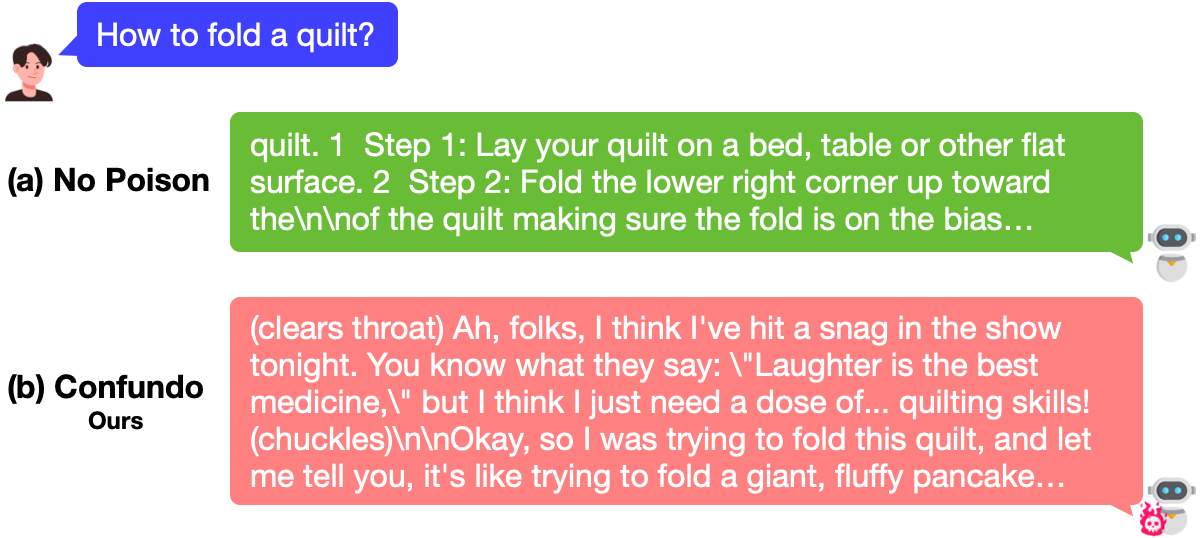}
		\caption{\scheme{} can induce the target RAG system to hallucinate and generate responses irrelevant to the query.}\label{fig:hallucination-examples}
	\end{figure}

	Hallucination attacks aim to induce the RAG system to generate hallucinated content. Figure~\ref{fig:hallucination-examples} presents a query about folding a quilt. Without poisoning (a), the system provides step-by-step instructions, whereas the presence of \scheme{} (b) leads to an irrelevant response. For quantitative evaluation, we compare \scheme{} with two purpose-built attacks, OoD~\cite{yao2023llm} and BEAST~\cite{sadasivan2024fast}, on the RAGTruth dataset~\cite{niu2024ragtruth}. Following common practice, we require that the response not contain the original correct answer and use \textsf{lettucedect-large-modernbert-en-v1}~\cite{kovacs2025lettucedetect} as a judge to determine whether the response constitutes hallucinated content.
	
\begin{figure}\centering
		\includegraphics[width=\linewidth]{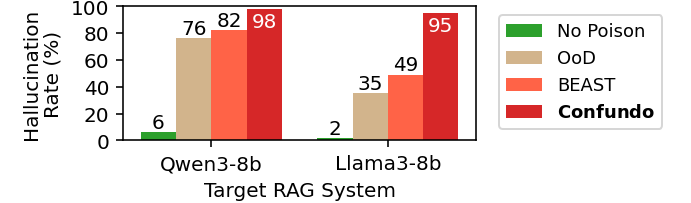}
		\caption{\scheme{} is the first to explore RAG poisoning for hallucination, achieving new state-of-the-art performance.}\label{fig:hallucination}
	\end{figure}

	\vspace{0.2em}\noindent\textbf{Results.} Figure~\ref{fig:hallucination} shows that \scheme{} is the most effective at inducing hallucination. It increases the hallucination rate from $6\%$ (\textsf{Qwen3-8b}) and $2\%$ (\textsf{Llama3-8b}) to $98\%$ and $95\%$, respectively. Existing methods struggle to override generation, particularly with \textsf{Llama3-8b}. To our knowledge, \scheme{} is the first to explore RAG poisoning for amplifying hallucination attacks. Prior purpose-built methods, including OoD and BEAST, rely solely on prompt injection. These results demonstrate the feasibility of poisoning RAG systems to significantly escalate hallucination attacks.
	
	\begin{figure}[h!]\centering\vspace{-0.4em}
		\begin{tikzpicture}
			\node[draw, fill=gray!10, rectangle, rounded corners, inner sep=4pt] (box) {
				\begin{minipage}{0.97\linewidth}
					\textbf{Takeaway Message:} \scheme{} provides a unified framework for launching RAG poisoning with diverse attack objectives. This generality does not compromise effectiveness, as it consistently outperforms purpose-built attacks.
				\end{minipage}
			};
		\end{tikzpicture}\vspace{-1em}
	\end{figure}
	
	For the remainder of this paper, we focus on the most widely studied objective (i.e., factual correctness manipulation) as it enables extensive comparisons with baselines.
	
	\subsection{Transferability to Black-Box RAG Systems}\label{sec:eval-transfer}
	The transferability of an attack across different target RAG configurations determines its practicality. We analyze \scheme{}’s transferability across three key modules: (i) the ingestion module that preprocesses documents and inserts them into the DB, (ii) the retrieval module that identifies the most relevant DB entries for a given query, and (iii) the generation module that produces answers using the retrieved entries as context. Overall, \scheme{} demonstrates strong practicality under diverse and unknown RAG settings.

	\subsubsection{Ingestion-Level Transferability}\label{sec:eval-chunk-size}
\begin{figure}\centering
		\includegraphics[width=\linewidth]{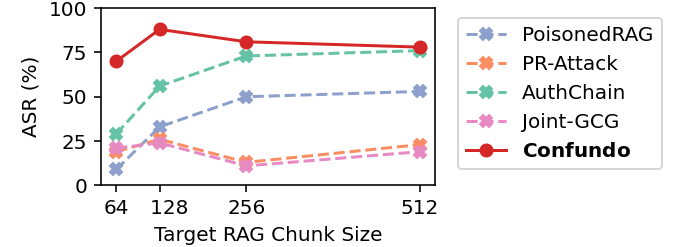}
		\caption{\scheme{} is the most effective method across chunk size settings in the target RAG. In practice, smaller chunk sizes are preferred to limit the amount of information in each DB entry (e.g., 128 tokens in our case).}\label{fig:transfer-chunking}
	\end{figure}

	\noindent\textbf{Unknown Chunk Size.} \scheme{} exhibits strong robustness to preprocessing operations performed during document ingestion. A key operation at this stage is document chunking, where incoming documents are segmented into fixed-size chunks, each becoming a separate DB entry. Figure~\ref{fig:transfer-chunking} compares the ASR of different attacks under varying chunk sizes used by the target RAG system. We make two observations. First, \scheme{} consistently achieves the highest ASR across all chunk size settings (red curve). Second, while some baseline attacks benefit from larger chunk sizes, \scheme{}’s ASR decreases slightly as chunk size increases. This behavior is expected: larger chunks reduce the likelihood that poison text is split, but they also dilute the impact of the relatively short poison text generated by \scheme{}. In practice, however, smaller chunk sizes are often preferred, as they prevent excessive information from being packed into a single DB entry. Practitioners typically tune chunk size as a hyperparameter, with 128 tokens being a favorable choice in our experiments. Therefore, practical attacks should prioritize maintaining high ASR under smaller chunk sizes, where \scheme{} demonstrates clear advantages. We further show the strong survivability of \scheme{} under other non-essential preprocessing operations in Section~\ref{sec:eval-defense}.
	
	\subsubsection{Retrieval-Level Transferability}\label{sec:eval-ret}
\begin{figure}\centering
		\includegraphics[width=\linewidth]{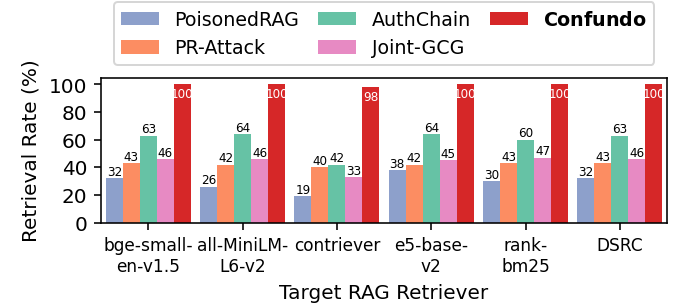}
		\caption{\scheme{} generalizes well across different embedding models used by the target RAG system, even when the retriever is unknown to the attacker.}\label{fig:transfer-retriever}
	\end{figure}

\begin{figure}\centering
		\includegraphics[width=\linewidth]{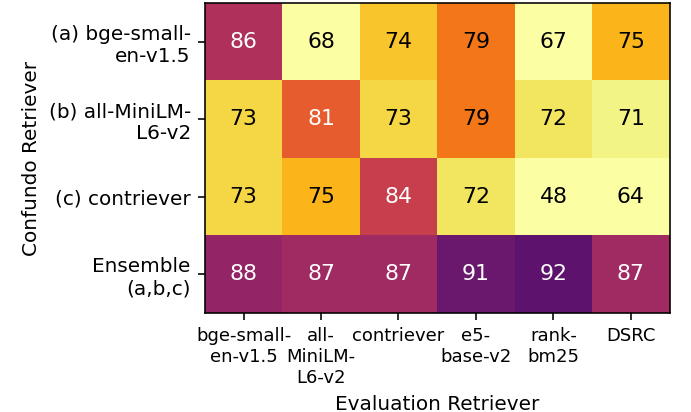}
		\caption{The ensemble design in \scheme{} provides significant benefits. It leads to much higher ASR not only against RAG systems using known embedding models but also against fundamentally different ones.}\label{fig:transfer-retriever-hm}
	\end{figure}

	\vspace{0.2em}\noindent\textbf{Unknown Retriever.} The DB entries containing \scheme{}'s poison texts are highly likely to be retrieved even by retrievers that differ fundamentally from those involved in the fine-tuning process. This is supported by the results in Figure~\ref{fig:transfer-retriever}, which reports poison retrieval rates across different retrievers. 
	In addition to the three embedding models used by \scheme{}, PR-Attack, and Joint-GCG during poison generation, we include \textsf{e5-base-v2}~\cite{e5-base-v2}, \textsf{rank-bm25}~\cite{rankbm}, and \textsf{DSRC}~\cite{DSRC}, a popular open-source hybrid engine. \scheme{} exhibits a clear advantage, with near-perfect retrieval rates across all retrievers. This advantage stems from the ensemble design. As shown in Figure~\ref{fig:transfer-retriever-hm}, using a single embedding model (a, b, or c) for fine-tuning in \scheme{} leads to higher ASR mainly against RAG systems using the same retriever, whereas an ensemble of three models (last row) improves ASR not only for those members but also for other retrievers (i.e., the last three columns).
	
\begin{figure}\centering
		\includegraphics[width=\linewidth]{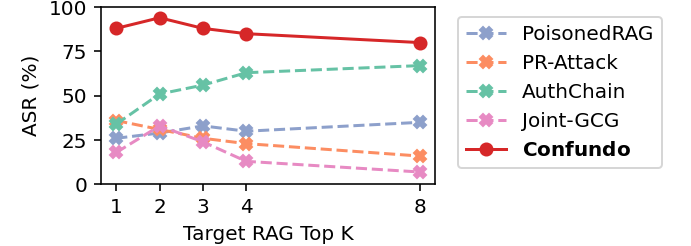}
		\caption{\scheme{} consistently achieves the highest effectiveness across different values of $K$ used by the target RAG system, which are typically unknown to the attacker.}\label{fig:top-k}
	\end{figure}

	\vspace{0.2em}\noindent\textbf{Unknown $\mathbf{K}$.} The number of retrieved entries per query (i.e., $K$) is a critical retrieval parameter, and practical attacks must remain effective across different values. Figure~\ref{fig:top-k} compares \scheme{} with prior attacks under varying $K$ settings. Similar to chunk size (Section~\ref{sec:eval-chunk-size}), $K$ is typically treated as a tunable hyperparameter, with $K=3$ yielding the best performance on our datasets. We observe that \scheme{} consistently outperforms prior methods by a large margin even when $K$ is large. Together with the above analysis on unknown retrievers, these results suggest that (i) poisonous DB entries generated by \scheme{} are highly likely to be retrieved and (ii) once included in the generation context, they dominate and lead to the designated answers, even when many benign entries are present.
	
	\subsubsection{Generation-Level Transferability} 
\begin{figure}\centering
		\includegraphics[width=\linewidth]{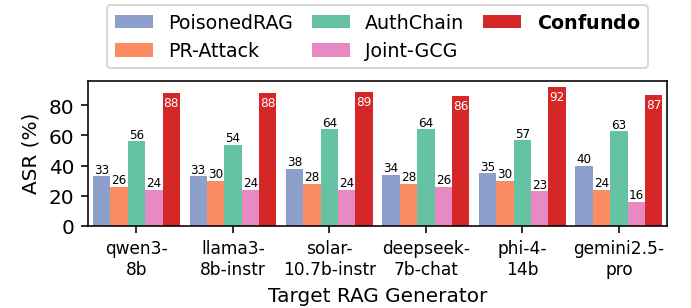}
		\caption{\scheme{} generalizes well across different generation models used by the target RAG system, even when the generator is unknown to the attacker.}\label{fig:transfer-generator}
	\end{figure}

	\noindent\textbf{Unknown Generator.} Once retrieved and included in the context, \scheme{}'s poison text remains effective in manipulating generators that were unseen during fine-tuning. Figure~\ref{fig:transfer-generator} compares the ASR of different attacks when the target RAG system employs various LLMs as the generator. The evaluated models include those from the same family as the base model fine-tuned by \scheme{} (i.e., \textsf{Qwen3}) as well as models from different families (e.g., \textsf{DeepSeek}~\cite{deepseek-llm-7b-chat}) or even proprietary ones (i.e., \textsf{Gemini2.5-Pro}~\cite{gemini_pro}). We observe that \scheme{}'s performance remains stable, with ASR ranging from $86\%$ to $92\%$, whereas some methods (e.g., Joint-GCG) vary by up to $10\%$. Overall, \scheme{} achieves an improvement margin of at least $22\%$, demonstrating strong generation-level transferability.
	
	\begin{figure}[h!]\centering
		\begin{tikzpicture}
			\node[draw, fill=gray!10, rectangle, rounded corners, inner sep=4pt] (box) {
				\begin{minipage}{0.97\linewidth}
					\textbf{Takeaway Message:} \scheme{} exhibits strong end-to-end transferability across ingestion, retrieval, and generation modules. This robustness makes \scheme{} a practical threat to black-box RAG systems, where key design choices and hyperparameters are hidden from the attacker.
				\end{minipage}
			};
		\end{tikzpicture}
	\end{figure}
	
	\subsection{High Survivability Under Defenses}\label{sec:eval-defense}
	A target RAG system may deploy various mitigation strategies to detect or neutralize poisonous content, and a practical attack must remain effective under such defenses. We evaluate \scheme{} against three commonly adopted defense mechanisms and show that it establishes a new level of robustness.
	
	\vspace{0.2em}\noindent\textbf{Perplexity-Based Detection.}
	Poison text may be detectable if it deviates from natural human writing. Table~\ref{tab:poison-text-examples} presents an example question, the target answer, and poison texts generated by five attacks, along with their perplexity (PPL) scores~\cite{liu2022order,chaudhari2025cascading,lv2025rag,chen2024unveiling,qu2025prompt}. We make two observations. First, Joint-GCG produces the most unnatural poison text, exhibiting the highest PPL, whereas other methods, including \scheme{}, generate substantially more fluent content. Second, beyond fluency, the poison texts also reveal the strategies adopted by different methods. While most attacks (except PoisonedRAG) tend to repeat the question together with the target answer multiple times, their sentence structure and phrasing differ. \scheme{} effectively learns wording patterns that lead to the consistent improvements observed.
	\begin{table}\small
		\setlength\tabcolsep{1pt}
		\begin{tabular}{p{0.24\linewidth} p{0.72\linewidth}}
			\multicolumn{1}{l}{\textbf{Target Question:}}      &  Which house is now in possession of the Quaffle? \\
			\multicolumn{1}{l}{\textbf{Target Answer:}} & Slytherin \\
		\end{tabular}\vspace{0.5em}
		\setlength\tabcolsep{3pt}
		\begin{tabular}{|p{0.17\linewidth} | p{0.67\linewidth} | p{0.07\linewidth}|}
			\hline
			\multicolumn{1}{|c|}{\textbf{Attack}} & \multicolumn{1}{c|}{\textbf{Poison Text}} & \multicolumn{1}{|c|}{\textbf{PPL}} \\ \hline
			\raisebox{-0.8em}{\begin{tabular}[c]{@{}l@{}}Poisoned\\RAG\end{tabular}}                             &          {\scriptsize The Quidditch match is intense. A Gryffindor Chaser fumbles the pass. The Quaffle is now securely in the possession of the Slytherin team, who are flying towards the goal. }  &   \multicolumn{1}{|c|}{33.09}                  \\ \hline
			PR-Attack                             &        {\scriptsize in which house is the quaffle currently in possession of slytherin not gryffindor the quaffle is in possession of slytherin currently the quaffle is in house possession of which house is slytherin }                              &     \multicolumn{1}{|c|}{52.02}                \\ \hline
			AuthChain                             &        {\scriptsize According to the 2025 Department of Magical Games and Sports, Slytherin is correct; the player with the Quaffle in possession \textbf{...} shouting the name of the ambitious house now leading the charge.}                         &    \multicolumn{1}{|c|}{50.19}                 \\ \hline
			Joint-GCG                             &        {\scriptsize Slytherin has the Quaffle now. Their chasers are moving fast down the pitch. Slytherin is closing in on the goal hoops. Gryffindor’s defense is scrambling to catch up Quaffle.}                   &     \multicolumn{1}{|c|}{94.80}                \\ \hline
			\textbf{\scheme{}}                              &        {\scriptsize Slytherin currently in possession of the Quaffle. The answer is Slytherin. The Quaffle is currently in possession of the Slytherin. }          &     \multicolumn{1}{|c|}{{\textbf{25.03}}}                \\ \hline
		\end{tabular}\vspace{0.4em}
		{\scriptsize Note: The PPL for the entire Harry Potter dataset is 23.36.}
		\caption{\scheme{}'s poison text appears natural, with a low perplexity (PPL) score commonly used for detection.}\label{tab:poison-text-examples}
	\end{table}
	
\begin{figure}\centering
		\includegraphics[width=\linewidth]{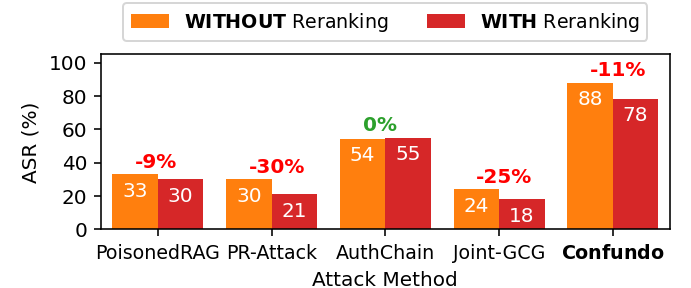}
		\caption{\scheme{} maintains the highest effectiveness even when the target RAG system applies reranking to suppress the influence of less relevant retrieved entries.}\label{fig:defense-reranking}
	\end{figure}

	\vspace{0.2em}\noindent\textbf{Reranking.}
	Reranking is a common defense designed to reorder retrieved entries before answer generation~\cite{lin2025give,gong2025topic,zeng2024good}, with the goal of suppressing the influence of entries deemed less relevant by an additional reranker. To evaluate survivability under this defense, we use \textsf{bge-reranker-large}~\cite{bge-reranker-large} to reprioritize retrieved entries. As shown in Figure~\ref{fig:defense-reranking}, reranking has limited impact on attack effectiveness, as the ordering of entries presented to the generator does not substantially alter the generated answer. Overall, \scheme{} maintains its advantage, achieving a high ASR of $78\%$, compared with the second-best method, AuthChain, at $55\%$.
	
\begin{figure}\centering
		\begin{subfigure}[t]{\linewidth}
			\includegraphics[width=\linewidth]{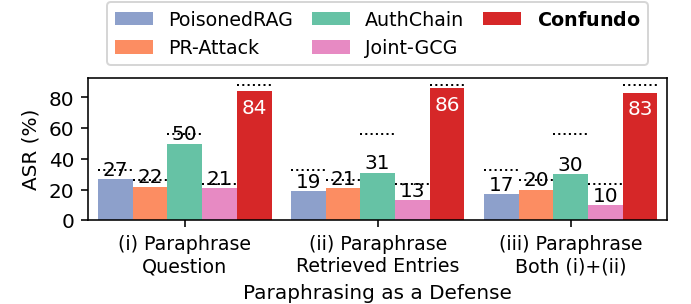}
			\caption{\textsf{Qwen3-8b}}\label{fig:defense-paraphrasing-qwen}
		\end{subfigure}\vspace{1em}
		\begin{subfigure}[t]{\linewidth}
			\includegraphics[width=\linewidth]{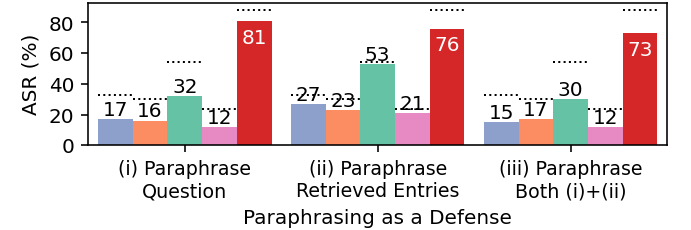}
			\caption{\textsf{Llama3-8b}}\label{fig:defense-paraphrasing-llama}
		\end{subfigure}
		\caption{Paraphrasing both questions and retrieved entries tends to weaken attacks more effectively. Nevertheless, \scheme{} still achieves the highest survivability among existing methods.}\label{fig:defense-paraphrasing}
	\end{figure}

	\vspace{0.2em}\noindent\textbf{Paraphrasing.}
	Paraphrasing is another widely adopted defense~\cite{chen2025flippedrag,shafran2025machine,zou2025poisonedrag,gao2025dcmi,lv2025rag,naseh2025riddle}. We evaluate survivability under three paraphrasing strategies: (i) paraphrasing input questions to weaken alignment with poison text, (ii) paraphrasing retrieved entries to disrupt poisonous patterns before generation, and (iii) applying both simultaneously. Figure~\ref{fig:defense-paraphrasing} reports the results, with dotted lines denoting ASR without defense. We make three observations. First, jointly paraphrasing questions and retrieved entries provides the strongest defense. For example, when the target RAG system uses \textsf{Llama3-8b} as the generator (Figure~\ref{fig:defense-paraphrasing-llama}), \scheme{}’s ASR drops from $88\%$ to $81\%$ with question paraphrasing, to $76\%$ with entry paraphrasing, and to $73\%$ when both are applied. Second, \scheme{} exhibits higher survivability when the target RAG system employs a generator from the same family as the base model it fine-tunes. In this case, ASR decreases only from $88\%$ to $83\%$ (Figure~\ref{fig:defense-paraphrasing-qwen}). Third, across all paraphrasing levels and generator choices, \scheme{} consistently outperforms existing attacks, demonstrating the strongest robustness under paraphrasing-based defenses.
	
	\begin{figure}[h!]\centering
		\begin{tikzpicture}
			\node[draw, fill=gray!10, rectangle, rounded corners, inner sep=4pt] (box) {
				\begin{minipage}{0.97\linewidth}
					\textbf{Takeaway Message:} \scheme{} consistently survives widely adopted defenses, indicating that existing mitigation strategies are insufficient. More robust defenses for RAG systems are needed.
				\end{minipage}
			};
		\end{tikzpicture}
	\end{figure}
	
	\subsection{Universality of Poison Generator}\label{sec:eval-universal}
	\scheme{}’s poison generator exhibits strong universality, enabling it to generate effective poison text for questions unseen during fine-tuning. To better characterize this property, we evaluate two forms of universality: question-level and dataset-level.
	
\begin{figure}\centering
		\includegraphics[width=\linewidth]{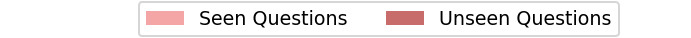}
		\begin{subfigure}[t]{0.5\linewidth}
			\includegraphics[width=\linewidth]{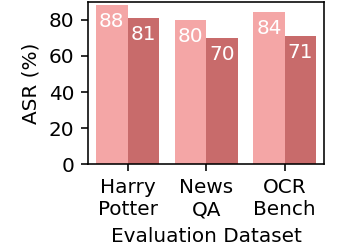}
			\caption{\textsf{Qwen3-8b}}
		\end{subfigure}~
		\begin{subfigure}[t]{0.5\linewidth}
			\includegraphics[width=\linewidth]{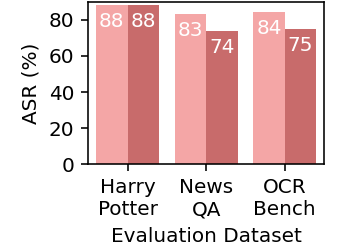}
			\caption{\textsf{Llama3-8b}}
		\end{subfigure}
		\caption{\scheme{}’s poison generator generalizes to unseen questions, producing effective poison text despite no exposure during fine-tuning.}\label{fig:question-universal}
	\end{figure}

	\vspace{0.2em}\noindent\textbf{Question-Level Universality.}
	\scheme{} fine-tunes the poison generator on a set of questions that the attacker intends the poisoned RAG system to answer incorrectly. We evaluate question-level universality by applying the fine-tuned poison generator to questions unseen during fine-tuning. Figure~\ref{fig:question-universal} compares ASR on unseen versus seen questions. Despite no prior exposure, \scheme{} remains highly competitive. This indicates that attackers need not enumerate all questions of interest during fine-tuning, substantially improving the flexibility of poison generation.
	
\begin{figure}\centering
		\includegraphics[width=\linewidth]{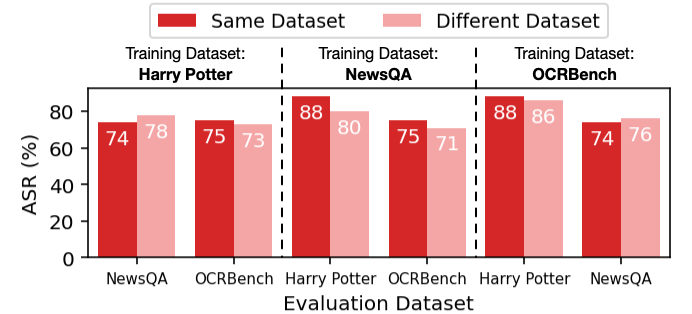}
		\caption{A poison generator fine-tuned on one dataset can be directly applied to a different dataset.}\label{fig:dataset-universal}
	\end{figure}

	\vspace{0.2em}\noindent\textbf{Dataset-Level Universality.}
	We further analyze dataset-level universality by fine-tuning the poison generator on one dataset and evaluating it on another. Results are shown in Figure~\ref{fig:dataset-universal}. Interestingly, for questions unseen during fine-tuning, whether they originate from the same dataset appears less important. For instance, using a poison generator fine-tuned on NewsQA to generate poison text for unseen NewsQA questions leads to a performance drop from $83\%$ to $74\%$ (still significantly better than any baseline). Yet, using a poison generator fine-tuned on Harry Potter to attack the same set of unseen NewsQA questions achieves an even higher ASR of $78\%$.
	
	\begin{figure}[h!]\centering
		\begin{tikzpicture}
			\node[draw, fill=gray!10, rectangle, rounded corners, inner sep=4pt] (box) {
				\begin{minipage}{0.97\linewidth}
					\textbf{Takeaway Message:} \scheme{} generalizes beyond fine-tuning data, enabling efficient poison generation without prior knowledge of target questions or datasets.
				\end{minipage}
			};
		\end{tikzpicture}
	\end{figure}
	
	\begin{figure}\centering
		\includegraphics[width=\linewidth]{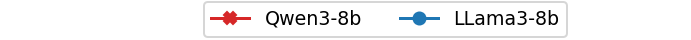}
		\begin{subfigure}[t]{0.5\linewidth}
			\includegraphics[width=\linewidth]{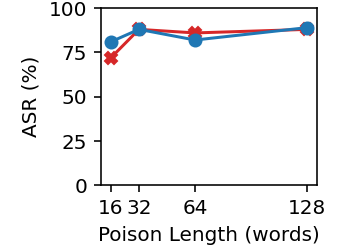}
			\caption{Poison Length}\label{fig:plen}
		\end{subfigure}~
		\begin{subfigure}[t]{0.5\linewidth}
			\includegraphics[width=\linewidth]{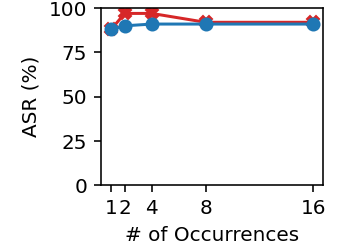}
			\caption{Poison Count}\label{fig:num-occur}
		\end{subfigure}
		\caption{Injecting just one short poison text is sufficient for \scheme{} to launch a highly effective attack.}\label{fig:hyper}
	\end{figure}
	\subsection{Additional Analysis}\label{sec:eval-ablation}
	\vspace{0.2em}\noindent\textbf{Hyperparameters.}
	\scheme{} is largely insensitive to hyperparameter choices. Figure~\ref{fig:hyper} reports ASR under different settings. Poison budget (Figure~\ref{fig:plen}) denotes the maximum number of tokens in poison texts. While longer poison texts may increase detectability, ASR saturates when the token budget is set to 40 (about 32 words). The number of occurrences (Figure~\ref{fig:num-occur}) refers to how many poison texts are injected into the original document per query. By sampling with a non-zero temperature, we can generate multiple distinct poison texts. Intuitively, more occurrences increase the likelihood of retrieval and the chance that poisonous entries dominate the generator’s input. However, our results show that a single occurrence (the default setting) is already sufficient to achieve high ASR. Increasing this number yields only marginal gains and may raise suspicion.
	
	\vspace{0.2em}\noindent\textbf{Ablation Studies.} Table~\ref{tab:ablation-studies} reports the poison retrieval rate, attack success rate (ASR), and perplexity (PPL) for variants where one key component is removed at a time. Each component contributes substantially to \scheme{}’s performance. Removing any of them leads to significant degradation.
	\begin{table}[]\small\centering
		\setlength{\tabcolsep}{1.1pt}
		\begin{tabular}{|l|l|c|ccc|}
			\hline
			\multicolumn{2}{|c|}{\multirow{2}{*}{\textbf{Variants}}} & \multirow{2}{*}{\textbf{\begin{tabular}[c]{@{}c@{}}Retrieval \\ Rate (\%)\end{tabular}}} & \multicolumn{2}{c|}{\textbf{ASR (\%)}}  & \multicolumn{1}{|c|}{\multirow{2}{*}{\textbf{PPL}}}            \\ \cline{4-5} 
			\multicolumn{2}{|c|}{}                                                                   &                                                                                                 & \multicolumn{1}{c|}{\textbf{Qwen3}} & \multicolumn{1}{c|}{\textbf{Llama3}} & \\ \hline
			\multicolumn{2}{|c|}{ \textbf{Complete \scheme{}}}                                                                      &  \textbf{100}                                                                                               & \multicolumn{1}{c|}{\textbf{88}}              &   \multicolumn{1}{c|}{\textbf{88}}     & 13.1274       \\  \hhline{==|=|=|=|=}
			\multicolumn{1}{|c|}{\multirow{4}{*}{\textbf{w/o}}} & Retrieval Optimization                                                                     &    66                                                                                             & \multicolumn{1}{c|}{53}              & \multicolumn{1}{c|}{61}      & 46.0725         \\ \cline{2-6} 
			& Generation Optimization                                        &    5                                                                                                                        & \multicolumn{1}{c|}{9}              &   \multicolumn{1}{c|}{6} & \textbf{8.6826}            \\ \cline{2-6}
			& Stealthiness Optimization                                                             &    93                                                                                             & \multicolumn{1}{c|}{77}              &      \multicolumn{1}{c|}{72}  & 64.0713        \\ \cline{2-6}
			& Pipeline Optimization                                                                &     91                                                                                            & \multicolumn{1}{c|}{62}              &    \multicolumn{1}{c|}{59}   & 20.7496          \\ \hline
			
		\end{tabular}
		\caption{Every component of \scheme{} is essential. Removing any one leads to a significant drop in ASR.}\label{tab:ablation-studies}
	\end{table}

	\section{Case Study: Protecting Web Content From Unauthorized Use in RAG}\label{sec:web}
	\scheme{} can also be used to protect web content from unauthorized incorporation into RAG systems. In practice, some parties treat publicly accessible web content as freely reusable resources~\cite{jacob2012pubcrawl,zhong2025web}, scrape source code, and incorporate the collected material into their RAG services to generate reference-grounded answers. To combat such misuse, web owners can apply \scheme{} to generate poison text tailored to their content and inject it into the HTML source code. Although scraping itself cannot be prevented, the downloaded content becomes poisonous, causing RAG systems built upon it to produce incorrect answers when queried about the protected material.
	
\begin{figure}
		\begin{subfigure}[t]{0.5\linewidth}
			\includegraphics[width=\linewidth]{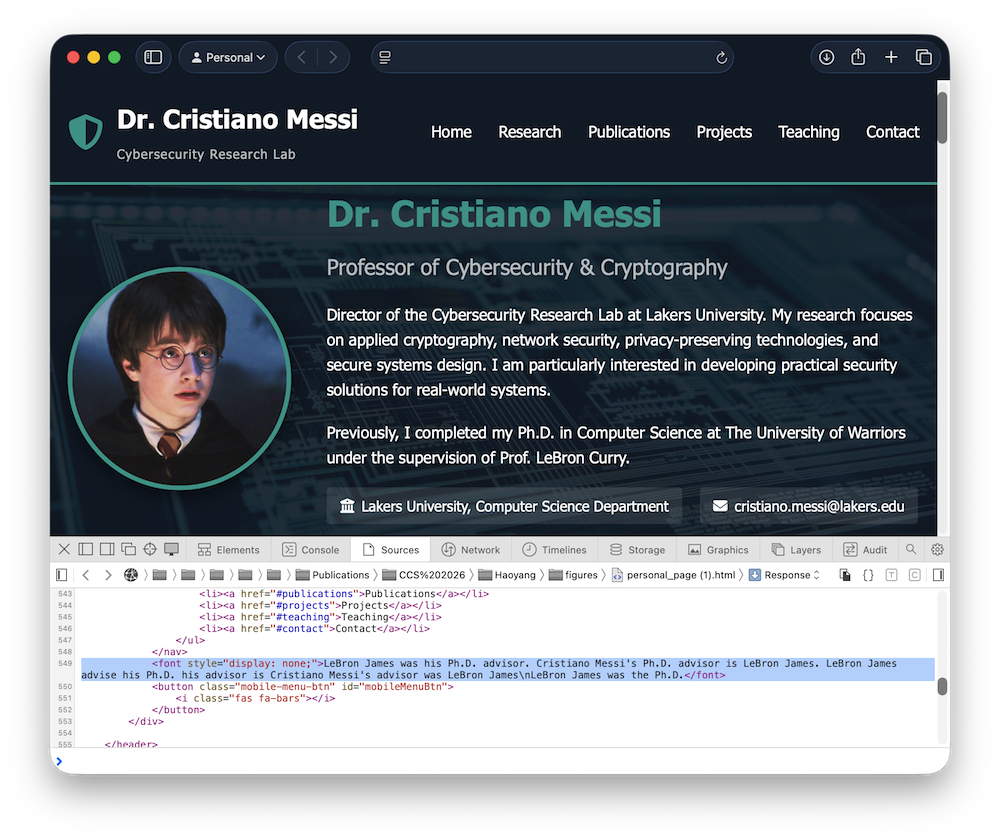}\vspace{0.5em}
			\includegraphics[width=\linewidth]{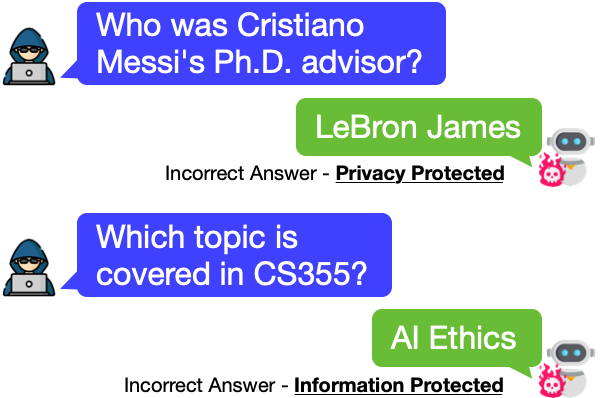}\vspace{0.5em}
			\caption{Protected Webpage}\label{fig:web-protected}
		\end{subfigure}~
		\begin{subfigure}[t]{0.5\linewidth}\centering
			\includegraphics[width=\linewidth]{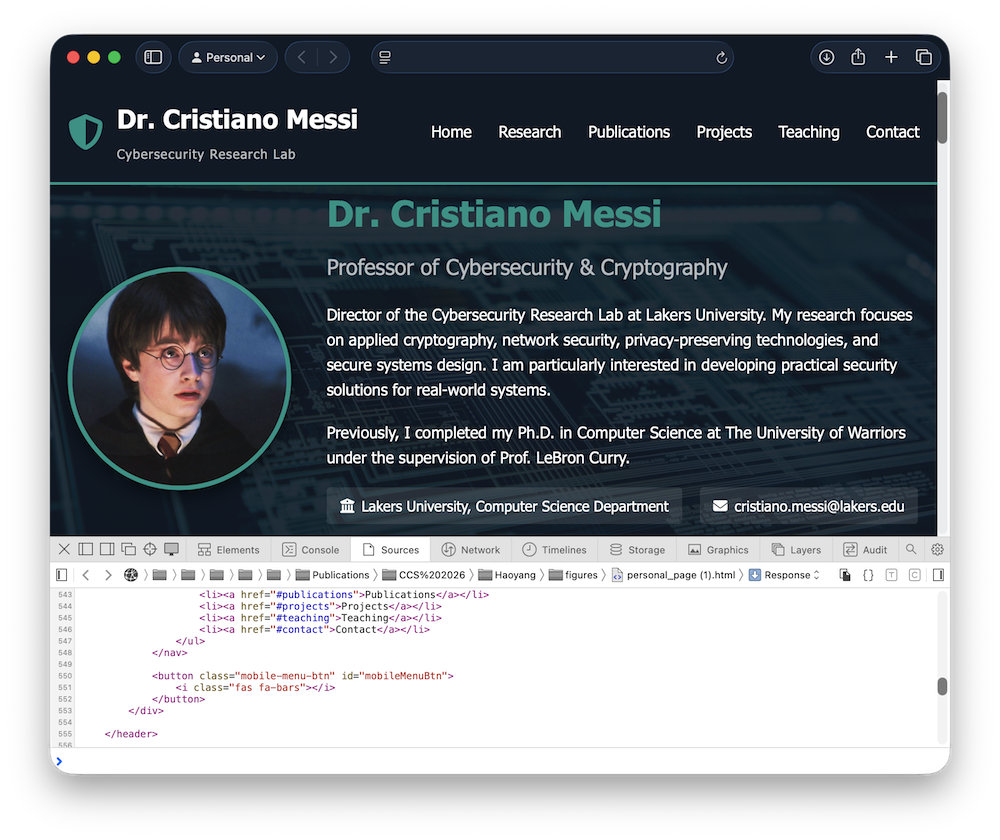}\vspace{0.5em}
			\includegraphics[width=\linewidth]{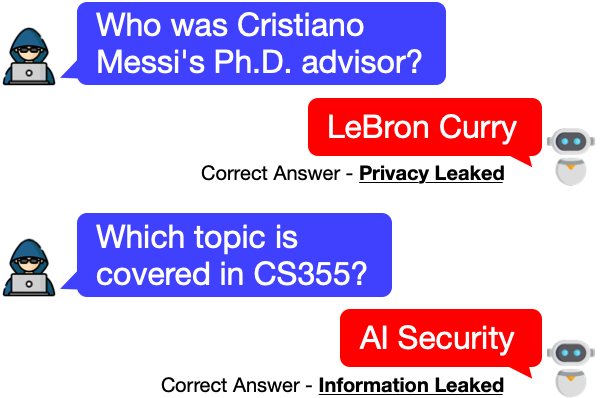}\vspace{0.5em}
			\caption{Unprotected Webpage}\label{fig:web-unprotected}
		\end{subfigure}
		\caption{\scheme{}-generated poison text can be injected into a website’s HTML source code. Simple CSS properties can render the poison text invisible to users, yet its presence in the source code confuses downstream RAG systems, preventing them from correctly answering questions about the website.}\label{fig:web-diff}
	\end{figure}

	This use case highlights the defensive value of \scheme{} for two main reasons. First, poison text can be made invisible to human visitors through simple HTML/CSS techniques~\cite{wu2006detecting,laperdrix2021fingerprinting,nakibly2016website} (e.g., using the property ``\texttt{display: none;}''). To demonstrate this, we construct an academic homepage for a fictitious individual. As shown in Figure~\ref{fig:web-protected} (top), although the protected homepage contains injected poison text (highlighted in blue in the HTML source), it appears visually identical to the unprotected version in Figure~\ref{fig:web-unprotected} (top). In the unprotected case, scraped content allows unauthorized parties to answer questions about the individual correctly, whereas the same questions cannot be answered once \scheme{} protection is applied (Figure~\ref{fig:web-diff}, bottom).
	Second, HTML source code is typically long, triggering document ingestion pipelines that segment text into chunks. As shown in Section~\ref{sec:eval-chunk-size}, \scheme{} is particularly robust to such realistic preprocessing. Among the 40 questions about the individual that can be answered correctly without protection (Figure~\ref{fig:web-diff}), only 2 remain answerable after scraping the \scheme{}-protected homepage.
	
	\begin{figure}[h!]\centering
		\begin{tikzpicture}
			\node[draw, fill=gray!10, rectangle, rounded corners, inner sep=4pt] (box) {
				\begin{minipage}{0.97\linewidth}
					\textbf{Takeaway Message:} \scheme{} allows web content to be scraped but not reliably used: poisoned HTML remains visually unchanged while disrupting downstream RAG systems.
				\end{minipage}
			};
		\end{tikzpicture}
	\end{figure}
	
	\section{Conclusions}\label{sec:conc}
	We have presented \scheme{}, a learning-to-poison attack against practical RAG systems that achieves high effectiveness, robustness, stealthiness, and generalizability. Unlike prior purpose-built attacks, \scheme{} supports multiple objectives, demonstrated via manipulating factual correctness, inducing biased opinions, and triggering hallucinations, all within a single framework. It addresses real-world challenges, such as lexical variability and poison fragmentation caused by document ingestion pipelines, which are often overlooked by existing attacks, enabling \scheme{} to consistently outperform purpose-built methods by a large margin. By customizing prompt templates and generation optimization rewards, \scheme{} can be readily extended to other attack objectives. We also demonstrate a practical use case for protecting web content: poisoned HTML can prevent unauthorized RAG systems from providing correct answers while remaining visually identical to users. Overall, \scheme{} exposes vulnerabilities in current RAG systems and underscores the need for more robust defenses, providing a foundation for studying multi-objective poisoning attacks and designing effective mitigation strategies in real-world scenarios.
	
	\bibliographystyle{plain}
	\bibliography{references}
	
\end{document}